\documentclass[letterpaper,twocolumn,english,pra; reprint]{revtex4-2}
\usepackage[T1]{fontenc}
\usepackage{color}
\usepackage{babel}
\usepackage{float}
\usepackage{mathrsfs}
\usepackage{amsmath}
\usepackage{amssymb}
\usepackage{graphicx}
\usepackage{xargs}[2008/03/08]
\usepackage[pdfusetitle,
 bookmarks=true,bookmarksnumbered=false,bookmarksopen=false,
 breaklinks=false,pdfborder={0 0 0},pdfborderstyle={},backref=false,colorlinks=true]
 {hyperref}

\makeatletter

\usepackage{babel}
\usepackage[justification=raggedright]{caption}

 {

      }

\usepackage{amsmath}

\allowdisplaybreaks[4]
\usepackage{orcidlink}

\usepackage{ragged2e}
\DeclareCaptionJustification{justified}{\justifying}
\ifdefined\showcaptionsetup
 
\fi

\ifdefined\showcaptionsetup
 \PassOptionsToPackage{caption=false}{subfig}
\fi
\usepackage{subfig}
\makeatother

\begin{document}
\title{Quantum-Limited Distance Estimation in Three-Dimensional Optical Superresolution}
\author{Junyan Li$^{1}$\orcidlink{0009-0008-1357-6590}}
\author{Shengshi Pang$^{1,2}$\orcidlink{0000-0002-6351-539X}}
\email{pangshsh@mail.sysu.edu.cn}

\affiliation{$^{1}$School of Physics, Sun Yat-sen University, Guangzhou, Guangdong
510275, China\\$^{2}$Hefei National Laboratory, University of Science
and Technology of China, Hefei 230088, China}
\begin{abstract}
Quantum superresolution reveals that the vanishing of separation sensitivity
in conventional imaging below the Rayleigh limit does not necessarily
indicate a fundamental loss of information in the optical field. However,
the quantum limit for estimating the physical distance between two
incoherent point sources in three-dimensional imaging systems and
its dependence on the spatial structure of the point-spread function
remains largely unknown. In this work, we derive the quantum-limited
precision for estimating the full distance between two incoherent
point sources with arbitrary intensity imbalance in a three-dimensional
spatially invariant imaging system. We show that the distance information
remains finite in the sub-Rayleigh regime and is governed by the second-order
displacement-response tensor of the point-spread function. The eigensystem
of this tensor determines the optimal relative orientation between
the two sources, and reflection symmetries of the point-spread function
can further provide a simplified means of identifying the optimal
orientation. This geometric structure is coordinate invariant and
provides a direct strategy for improving resolution by physically
rotating an anisotropic imaging system to align its optimal principal
response direction with the source displacement. For a general three-dimensional
Gaussian point-spread function, the response tensor is proportional
to the inverse spatial covariance, establishing a direct connection
between quantum-limited distance precision and the geometry of Gaussian
distribution.
\end{abstract}
\maketitle
\global\long\def\hro{\hat{\rho}}%
\global\long\def\bra#1{\langle#1|}%
\global\long\def\ket#1{|#1\rangle}%
\newcommandx\ksi[1][usedefault, addprefix=\global, 1=]{\ket{\psi_{#1}}}%
\newcommandx\bsi[1][usedefault, addprefix=\global, 1=]{\bra{\psi_{#1}}}%
\newcommandx\xk[1][usedefault, addprefix=\global, 1=k]{X_{#1}}%
\newcommandx\yk[1][usedefault, addprefix=\global, 1=k]{Y_{#1}}%
\newcommandx\zk[1][usedefault, addprefix=\global, 1=k]{Z_{#1}}%
\global\long\def\dx{d_{x}}%
\global\long\def\dy{d_{y}}%
\global\long\def\dz{d_{z}}%
\global\long\def\xb{\bar{X}}%
\global\long\def\yb{\bar{Y}}%
\global\long\def\zb{\bar{Z}}%
\newcommandx\hp[2][usedefault, addprefix=\global, 1=, 2=]{\hat{P}^{#2}_{#1}}%
\newcommandx\avg[4][usedefault, addprefix=\global, 1=, 2=, 3=, 4=]{\langle\hp[#1][#4]\rangle^{#3}_{#2}}%
\global\long\def\qo{\langle\psi_{1}|\psi_{2}\rangle}%
\newcommandx\av[1][usedefault, addprefix=\global, 1=]{\langle#1\rangle}%
\global\long\def\norm#1{\left|#1\right|^{2}}%
\newcommandx\nei[2][usedefault, addprefix=\global, 1=]{e^{-i#2\hp[#1]}}%
\newcommandx\nk[1][usedefault, addprefix=\global, 1=k]{N_{#1}}%
\newcommandx\px[1][usedefault, addprefix=\global, 1=x]{\partial_{#1}}%
\newcommandx\py[1][usedefault, addprefix=\global, 1=y]{\partial_{#1}}%
\newcommandx\pz[1][usedefault, addprefix=\global, 1=z]{\partial_{#1}}%
\newcommandx\nt[1][usedefault, addprefix=\global, 1=]{N^{#1}_{{\rm tot}}}%
\newcommandx\nd[1][usedefault, addprefix=\global, 1=]{N^{#1}_{{\rm diff}}}%
\newcommandx\hl[2][usedefault, addprefix=\global, 1=]{\hat{\mathcal{L}}^{#1}_{#2}}%
\newcommandx\gi[1][usedefault, addprefix=\global, 1=i]{g_{#1}}%
\global\long\def\var{{\rm Var}}%
\newcommandx\hr[1][usedefault, addprefix=\global, 1=]{\mathcal{H}^{#1}_{r}}%
\newcommandx\ha[1][usedefault, addprefix=\global, 1=]{\mathcal{H}^{#1}_{\alpha}}%

\global\long\def\hrn{\mathcal{H}^{({\rm num})}_{r}}%
\global\long\def\hrd{\mathcal{H}^{({\rm den})}_{r}}%
\newcommandx\ei[1][usedefault, addprefix=\global, 1=]{e^{#1}}%
\global\long\def\dlm#1{\left.#1\right|_{r\rightarrow0}}%
\global\long\def\hrop{\mathcal{H}^{({\rm opt})}_{r}}%
\global\long\def\hroc{\mathcal{H}^{({\rm cir})}_{r}}%

\global\long\def\ap{\alpha_{{\rm opt}}}%
\global\long\def\aw{\alpha_{{\rm wor}}}%

\newcommandx\et[2][usedefault, addprefix=\global, 1=, 2=]{\eta^{#2}_{#1}}%
\newcommandx\kp[2][usedefault, addprefix=\global, 1=, 2=]{\kappa^{#2}_{#1}}%

\section{Introduction}

Optical imaging can be viewed as a parameter-encoding process: spatial
information of an object is mapped into an optical field by the imaging
system and subsequently into measurement statistics by a given detection
scheme \citep{Goodman1996}. The conventional benchmark for the performance
of optical imaging is provided by the Rayleigh criterion \citep{Rayleigh1880},
where direct intensity measurements lose sensitivity to small separations
as two optical sources increasingly overlap. Nevertheless, Tsang \textit{et
al}. \citep{Tsang2016} showed that the loss of resolution in direct
imaging does not imply a fundamental loss of separation information
from the field: the quantum Cram\'er-Rao bound (QCRB) \citep{Cramer2016,Tsang2011},
the ultimate precision limit given by the quantum estimation theory,
can remain finite for vanishing separation by accessing appropriate
spatial modes. This result established the principle of quantum superresolution
and revealed the importance of exploiting the full spatial structure
of an optical field in resolving two sources.

Quantum superresolution has since been explored through image inversion
interferometry \citep{Tang2016}, heterodyne detection \citep{Yang2016},
edge-coherence inversion \citep{Tham2017}, and two-photon interference
\citep{Parniak2018,Thachil2023}, etc. A growing number of experimental
implementations have also been reported \citep{Tang2016,Hassett2018,Zhou2019a,Datta2021a,Zanforlin2022,Gorecki2022}.
To establish the ultimate limits of superresolution under realistic
conditions, extensive theoretical studies have been carried out on
more general source models, including thermal sources \citep{Lupo2016,Nair2016,Wang2021},
unbalanced-intensity sources \citep{Rehacek2017,Rehacek2018,Li2024},
as well as partially coherent and coherent optical sources \citep{Larson2018,Tsang2019c,Larson2019a,De2021a,Wadood2021,Karuseichyk2022,Liang2023}.
More general imaging scenarios have also been systematically investigated
\citep{Zhou2019,Tsang2019,Tsang2019b,Bisketzi2019,Gorecki2022a,Tsang2023,Zhou2024}.

In realistic optical systems, the point-spread function is intrinsically
a multidimensional response that determines how spatial displacements
of the object are encoded into the optical field. Quantum superresolution
has therefore been extended beyond one-dimensional source separation
to multidimensional imaging configurations \citep{Ang2017,Wang2021a,Prasad2020a,Gosalia2024,Prasad2020b,tzyy-cwb4,Yu2018,Prasad2020}.
These studies show that the ultimate estimation precision of a spatial
parameter is determined by the interplay between the geometry of the
source configurations, the response properties of the imaging system,
suggesting the imaging system itself can be engineered to improve
the precision of a target estimation task \citep{Paur2018,Paur2019,Xin2021}.
In three-dimension imaging, this interplay becomes particularly nontrivial
because displacements along different spatial directions may experience
different responses. A general imaging system can therefore possess
multiple principal directions with generally unequal displacement-response
sensitivities. The resulting anisotropy determines how estimation
precision varies with the orientation of the source displacement and
must be taken into account when optimizing three-dimensional imaging
performance.

In this work, we derive the quantum limit for estimating the physical
distance between two incoherent point sources in a three-dimensional
imaging system with arbitrary intensity imbalance. We show that the
information of the distance between two sources remains finite in
the sub-Rayleigh regime, even when the two sources approach spatial
coincidence. The resulting resolution is governed by the second-order
spatial-frequency moment tensor of the point-spread function, whose
eigensystem identifies the principal displacement-response directions
and provides the optimal and worst relative orientation between the
two sources. The orientational dependence of the distance precision
suggests an interesting approach for superresolution optimization:
for an anisotropic point-spread function, even with a fixed source
configuration, the imaging system can be rotated to align the source
displacement direction with the principal response direction determined
by the second-order spatial-frequency response tensor of the point-spread
function, thereby optimizing the attainable resolution of the two
sources. We further show that the reflection symmetries of the point-spread
function can directly identify the principal directions of the displacement
response: when the point-spread function possesses reflection planes,
the corresponding normal directions become principal response directions.
A single reflection plane separates one principal direction, whereas
multiple orthogonal reflection planes determine a complete principal
response basis. Finally, a general three-dimensional Gaussian point-spread
function is considered as an example to illustrate the above results.

\section{Preliminaries}

Consider two incoherent optical point sources in a spatially invariant
three-dimensional imaging system, located at $\boldsymbol{s_{1}}=\left(X_{1},Y_{1},Z_{1}\right)^{\intercal}$
and $\boldsymbol{s_{2}}=\left(X_{2},Y_{2},Z_{2}\right)^{\intercal}$,
and separated by a distance $r=\left|\boldsymbol{s_{2}}-\boldsymbol{s_{1}}\right|$.
Assume that the probability of detecting more than one photon within
the coherence time is negligible and photon arrivals with the mean
photon numbers $N_{1}$ and $N_{2}$ from each source during the measurement
interval. The total average photon number is therefore $\nt=N_{1}+N_{2}$,
and the normalized intensity imbalance can be defined as $\epsilon=\frac{N_{2}-N_{1}}{N_{2}+N_{1}}\in[-1,1]$.
The average state of a detected photon can then be described by the
density operator \citep{Mandel1959},
\begin{equation}
\begin{aligned}\hro= & \sum^{2}_{i=1}q_{i}\ei[-i\boldsymbol{s_{i}}.\hat{\boldsymbol{P}}]\ksi\bsi\ei[i\boldsymbol{s_{i}}.\hat{\boldsymbol{P}}]\end{aligned}
,\label{eq:rho}
\end{equation}
where $q_{1,2}=\frac{1\mp\epsilon}{2}$ and $\hat{\boldsymbol{P}}=(\hp[x],\hp[y],\hp[z])^{\intercal}=-i\boldsymbol{\nabla}$
denotes the three-dimensional momentum operator. The state $\ksi$
is defined as
\begin{equation}
\ksi=\int^{\infty}_{-\infty}d^{3}\boldsymbol{s}\psi(\boldsymbol{s})\ket{\boldsymbol{s}},
\end{equation}
where $\boldsymbol{s}=(x,y,z)^{\intercal}$ and $\psi(\boldsymbol{s})$
is the point-spread function of the imaging system, assumed to be
real and inversion-symmetric, i.e., $\psi(\boldsymbol{s})=\psi(-\boldsymbol{s})$
\citep{Goodman1996}, throughout this work.

\begin{figure}[!h]
\centering{}\includegraphics[width=0.85\columnwidth]{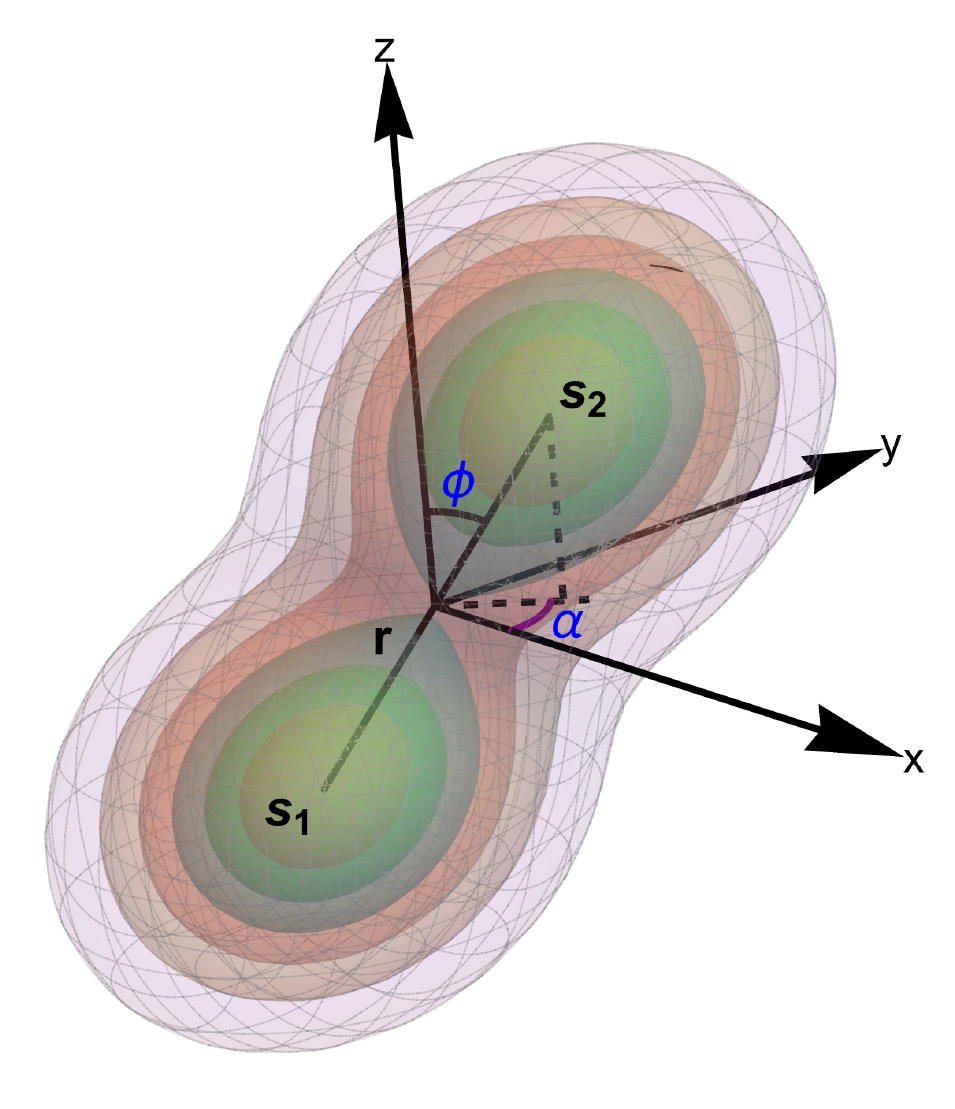}\caption{\label{fig:model}Schematic illustration of two incoherent optical
point sources located at positions $\boldsymbol{s_{1}}$ and $\boldsymbol{s_{2}}$
in three-dimensional space. The source separation is characterized
by the distance $r=\left|\boldsymbol{s_{2}}-\boldsymbol{s_{1}}\right|$
and the displacement direction $\boldsymbol{u}=(\boldsymbol{s_{2}}-\boldsymbol{s_{1}})/r$,
which is parameterized by the azimuthal angle $\alpha$ and the polar
angle $\phi$.}
\end{figure}

It is convenient to decompose their coordinates into the centroid
$\bar{\boldsymbol{s}}=\frac{\boldsymbol{s_{1}}+\boldsymbol{s_{2}}}{2}$
and the displacement $\boldsymbol{d}=\boldsymbol{s_{2}}-\boldsymbol{s_{1}}=(d_{x},d_{y},d_{z})^{\intercal}$.
The full distance is $r=\left|\boldsymbol{d}\right|$, with its orientation
specified by the normalized vector $\boldsymbol{u}=\mathit{\boldsymbol{d}}/r$.
The source positions can then be written as
\begin{equation}
\boldsymbol{s_{1,2}}=\bar{\boldsymbol{s}}\mp\frac{r}{2}\mathit{\boldsymbol{u}}.
\end{equation}
In Fig. \ref{fig:model}, the unit displacement vector $\boldsymbol{u}$
is parameterized as $\boldsymbol{u}=\left(\cos\alpha\sin\phi,\sin\alpha\sin\phi,\cos\phi\right)^{\top}$,
where $\alpha$ and $\phi$ denote the azimuthal and polar angles
specifying the orientation of the displacement vector. The plane orthogonal
to $\boldsymbol{u}$ can then be spanned by $\boldsymbol{v_{\alpha}}=\frac{1}{\sin\phi}\frac{\partial\boldsymbol{u}}{\partial\alpha}=\left(-\sin\alpha,\cos\alpha,0\right)^{\top}$
and $\boldsymbol{v_{\phi}}=\frac{\partial\boldsymbol{u}}{\partial\phi}=\left(\cos\alpha\cos\phi,\sin\alpha\cos\phi,-\sin\phi\right)^{\top}$.

\section{Quantum Superresolution of Source distance}

\subsection{Estimation precision of full distance}

In the imaging model introduced above, the source coordinates are
encoded into the spatial state of each detected photon through translations
of the point-spread function. Resolving the two sources can therefore
be formulated as estimating these encoded spatial parameters from
optical measurement outcomes. Since any physically admissible measurement
is described by a positive-operator-valued measure, quantum estimation
theory provides a measurement-independent bound on the Fisher information
attainable over all such measurements. Within the framework of quantum
multiparameter estimation, we first specify the \textit{relative}
displacement by its Cartesian components $(d_{x},d_{y},d_{z})^{\intercal}$,
the unknown parameters to estimate in this case can be collected as
the centroid position $(\xb,\yb,\zb)^{\top}$ and the displacement
vector $(d_{x},d_{y},d_{z})^{\top}$. Although the single-photon density
operator $\hro$ in Eq. \eqref{eq:rho} acts on an infinite-dimensional
Hilbert space, it has rank at most two. Its QFIM can therefore be
evaluated using its two nonzero eigenvalues and the parameter derivatives
of the corresponding eigenvectors $\hro$ \citep{Liu2016}. The resulting
Cartesian QFIM is given in Eq. \eqref{eq:fimc}, with the derivation
provided in Appendix \ref{subsec:com3}. According to the quantum
Cram\'er-Rao bound, the estimation precision of each displacement
component $d_{i},i=x,y,z$ is determined by the diagonal elements
of the inverse QFIM. In the limit $\boldsymbol{d}\rightarrow0$, the
effective quantum information for each displacement component, with
the remaining parameters treated as nuisance parameters, is
\begin{equation}
\begin{aligned}\left.\mathcal{H}_{d_{i}}\right|_{d_{i}\to0}= & \nt(1-\epsilon^{2})/(\boldsymbol{K}^{-1})_{ii},\,i=1,2,3.\end{aligned}
\label{eq:comhr}
\end{equation}
Here, $\boldsymbol{K}$ describes the local response of the point-spread-function
amplitude to infinitesimal spatial displacements
\begin{equation}
\begin{aligned}{\bf K}= & \int d^{3}\boldsymbol{s}\,[\boldsymbol{\nabla}\psi(\boldsymbol{s})][\boldsymbol{\nabla}\psi(\boldsymbol{s})]^{\mathsf{T}}\\
= & \left(\begin{array}{ccc}
\langle\partial_{x}\psi|\partial_{x}\psi\rangle & \langle\partial_{x}\psi|\partial_{y}\psi\rangle & \langle\partial_{x}\psi|\partial_{z}\psi\rangle\\
\langle\partial_{y}\psi|\partial_{x}\psi\rangle & \langle\partial_{y}\psi|\partial_{y}\psi\rangle & \langle\partial_{y}\psi|\partial_{z}\psi\rangle\\
\langle\partial_{z}\psi|\partial_{x}\psi\rangle & \langle\partial_{z}\psi|\partial_{y}\psi\rangle & \langle\partial_{z}\psi|\partial_{z}\psi\rangle
\end{array}\right).
\end{aligned}
\label{eq:k}
\end{equation}
Thus, $\boldsymbol{K}$ is a real symmetric positive-semidefinite
tensor determined solely by the spatial structure of the point-spread
function. In the sub-Rayleigh regime, the component precisions \eqref{eq:comhr}
are set by the derivative structure of the point-spread function and
remain finite as $r\to0$. The Cartesian displacement components therefore
do not suffer from Rayleigh's curse at the quantum limit.

In the multiparameter estimation, the quantum Cram\'er-Rao bound
is not generally attainable because the optimal measurements for different
parameters may be incompatible. For the real point-spread function
considered here, however, $\hro$ and its symmetric logarithmic derivatives
admit real symmetric representations in the position basis. Their
commutator is therefore real antisymmetric, and hence 
\begin{equation}
{\rm Tr}(\hro[\hl{g_{i}},\hl{g_{j}}])={\rm ImTr}(\hro\hl{g_{i}}\hl{g_{j}})=0.
\end{equation}
Consequently, the quantum Cram\'er-Rao bound coincides with the asymptotically
attainable Holevo bound and provides an achievable precision limit
for the distance-estimation problem considered here \citep{Ragy2016,Rehacek2017}.

While the above Cartesian-component analysis provides a generalization
from the one-dimensional quantum superresolution, the quantity of
primary physical interest that determines the resolution of higher-dimensional
imaging systems is the full distance $r$ between two optical sources.
We therefore reparametrize the \textit{relative} displacement as $\mathit{\boldsymbol{d}}=r\boldsymbol{u}$,
where $\boldsymbol{u}$ can be parameterized by the azimuthal $\alpha$
and polar angles $\phi$. The density operator $\hro$ then depends
on the parameter set $\boldsymbol{g}=(\xb,\yb,\zb,r,\alpha,\phi)^{\top}$.
By diagonalizing $\hro$, the QFIM with respect to the unknown parameter
vector $\boldsymbol{g}$ can be obtained straightforwardly by its
definition. The quantum-limited precision for estimating an arbitrary
distance $r$ is derived in Appendix \ref{subsec:3dqf}.

In the limit of $r\rightarrow0$, the quantum-limited precision for
the distance $r$ can be simplified to
\begin{equation}
\left.\hr\right|_{r\to0}=\frac{\nt\left(1-\epsilon^{2}\right)}{\boldsymbol{u}^{\intercal}\boldsymbol{K}^{-1}\boldsymbol{u}}.\label{eq:hr}
\end{equation}
It shows that the distance between the two sources remains estimable
with finite precision even when the two sources approach coincidence.
It is also worth emphasizing that the angular variables $\alpha$
and $\phi$ enter Eq. \eqref{eq:hr} only through the unit vector
$\boldsymbol{u}$. Their numerical values depend on the chosen coordinate
axes, whereas the scalar $\boldsymbol{u}^{\intercal}\boldsymbol{K}^{-1}\boldsymbol{u}$
is invariant under a change of coordinates, provided that both $\boldsymbol{u}$
and the matrix representation of $\boldsymbol{K}$ are transformed
consistently. Hence, the distance-estimation precision is independent
of the coordinate system, as shown explicitly in Appendix \ref{subsec:Rotation-invariance}.

When the displacement and the imaging response are restricted to
a two-dimensional subspace, this expression is reduced to
\begin{equation}
\begin{aligned}\left.\hr[\text{2D}]\right|_{r\to0}= & \frac{\nt\left(1-\epsilon^{2}\right)}{\boldsymbol{u}^{\intercal}_{\perp}\boldsymbol{K}^{-1}_{\perp}\boldsymbol{u}_{\perp}}\\
= & \frac{\nt\left(1-\epsilon^{2}\right)\left(\avg[x][][][2]\avg[y][][][2]-\av[{\hp[x]\hp[y]}]^{2}\right)}{\avg[x][][][2]\sin^{2}\alpha+\avg[y][][][2]\cos^{2}\alpha-\av[{\hp[x]\hp[y]}]\sin(2\alpha)}.
\end{aligned}
\end{equation}
.

\subsection{Orientational enhancement\label{subsec:Orientational-enhancement}}

Eq. \eqref{eq:hr} shows that the estimation precision is determined
by the displacement direction $\boldsymbol{u}$ and the point-spread-function
response tensor $\boldsymbol{K}$. For an anisotropic point-spread-function
response, different \textit{\emph{displacement}} orientations between
the sources can therefore lead to different distance precisions.

To determine the optimal estimation precision, we consider
\begin{equation}
f(\boldsymbol{u})=\boldsymbol{u}^{\intercal}\boldsymbol{K}^{-1}\boldsymbol{u},\,{\rm subject\,to}\;\boldsymbol{u}^{\top}\boldsymbol{u}=1.
\end{equation}
Since $\hr$ \eqref{eq:hr} is inversely proportional to $f(\boldsymbol{u})$,
minimizing $f$ leads to the optimal distance precision, whereas maximizing
$f$ gives the least favorable precision. Introducing a Lagrange multiplier
$\mu$, we define
\begin{equation}
\mathcal{L}(\boldsymbol{u},\mu)=\boldsymbol{u}^{\intercal}\boldsymbol{K}^{-1}\boldsymbol{u}-\mu\left(\boldsymbol{u}^{\top}\boldsymbol{u}-1\right).
\end{equation}
The stationary condition is
\begin{equation}
\boldsymbol{\nabla}_{\boldsymbol{u}}\mathcal{L}=2\boldsymbol{K}^{-1}\boldsymbol{u}-2\mu\boldsymbol{u}=0,
\end{equation}
which leads to
\begin{equation}
\boldsymbol{K}^{-1}\boldsymbol{u}=\mu\boldsymbol{u}.
\end{equation}
Thus, the stationary directions are the eigenvectors of $\boldsymbol{K}^{-1}$,
or equivalently those of $\boldsymbol{K}$. Denoting the eigenvalues
of $\boldsymbol{K}$ by $\lambda_{i}$, an eigenvector with eigenvalue
$\lambda_{i}$ gives $\boldsymbol{u}^{\intercal}\boldsymbol{K}^{-1}\boldsymbol{u}=\frac{1}{\lambda_{i}}$.
$\lambda_{i}$ must be non-negative, as $\boldsymbol{K}$ is positive
semidefinite according Eq. \eqref{eq:k}.

The maximal precision is then achieved when $\boldsymbol{u}$ is in
parallel with the eigenvector associated with the largest eigenvalue
$\lambda_{\max}$,
\begin{equation}
\hr[{\rm opt}]=\nt\left(1-\epsilon^{2}\right)\lambda_{\max},\label{eq:ophr}
\end{equation}
while the minimum precision occurs with the smallest eigenvalue $\lambda_{\min}$,
\begin{equation}
\hr[{\rm wor}]=\nt\left(1-\epsilon^{2}\right)\lambda_{\min}.\label{eq:whr}
\end{equation}
The maximum precision gain achievable by the optimal alignment is
therefore
\begin{equation}
\xi=\frac{\lambda_{\max}}{\lambda_{\min}}.
\end{equation}

This result implies that a stronger anisotropy of the point-spread-function
response leads to a larger increase ratio in the distance-estimation
precision. The optimization above is expressed in terms of $\boldsymbol{u}$,
which compactly describes how the source-displacement direction probes
the principal response axes of the imaging system.

Inspired by the dependence of the precision $\hr$ on the relative
orientation $\boldsymbol{u}$ between the two point sources, one can
in principle improve the precision by rotating one point source around
the other to optimize $\boldsymbol{u}$. But such an approach could
be constrained by the intrinsic physical origin of the point sources
and is not always feasible, so a more accessible approach would be
rotating the imaging system rather than the sources. Rotating the
imaging system effectively rotates the point-spread function $\psi(\boldsymbol{s})$
while leaving $\boldsymbol{u}$ unchanged, which is equivalent to
rotating $\boldsymbol{u}$ with $\psi(\boldsymbol{s})$ fixed, so
it can be used to optimize the precision $\hr$. Below, we prove this
equivalence in detail.

Suppose we rotate the imaging system by a real unitary matrix $R$,
i.e., the point-spread function becomes
\begin{equation}
\psi_{R}(\boldsymbol{s})=\psi(R^{-1}\boldsymbol{s}).
\end{equation}
It can be verified the gradient of $\psi(\boldsymbol{s})$ is changed
to $\boldsymbol{\nabla}\psi_{R}(\boldsymbol{s})=R(\boldsymbol{\nabla}\psi)(R^{-1}\boldsymbol{s})$,
and therefore the response tensor $\boldsymbol{K}$ \eqref{eq:k}
becomes
\begin{equation}
\boldsymbol{K}_{R}=R\boldsymbol{K}R^{\mathsf{T}}.
\end{equation}
Thus, the precision $\hr$ \eqref{eq:hr} is changed to
\begin{equation}
\left.\hr\right|_{r\to0}=\frac{\nt\left(1-\epsilon^{2}\right)}{\boldsymbol{u}^{\intercal}R\boldsymbol{K}^{-1}R^{\mathsf{T}}\boldsymbol{u}},\label{eq:hr-3}
\end{equation}
where $R^{\mathsf{T}}=R^{-1}$ has been used as $R$ is real and unitary.

\begin{figure}[!t]
\begin{centering}
\subfloat[\centering]{\includegraphics[width=0.5\columnwidth]{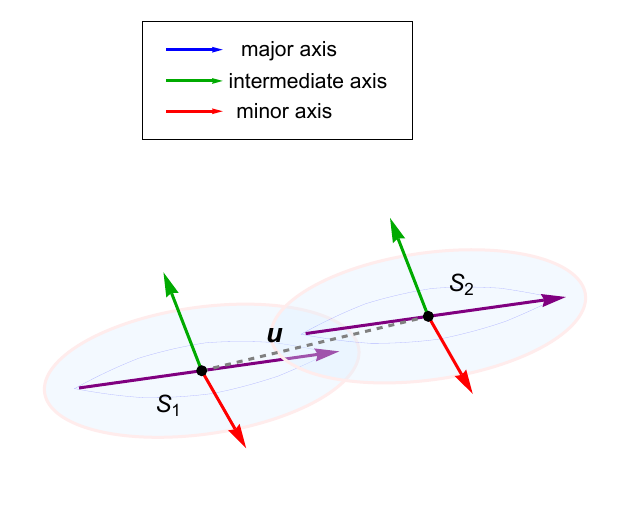}

} \subfloat[\centering]{\includegraphics[width=0.5\columnwidth]{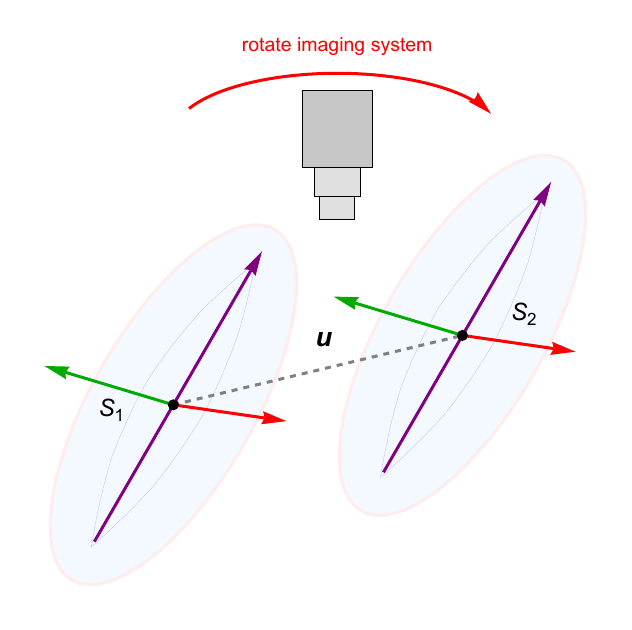}

}
\par\end{centering}
\begin{centering}
\subfloat[\centering]{\begin{centering}
\includegraphics[width=0.5\columnwidth]{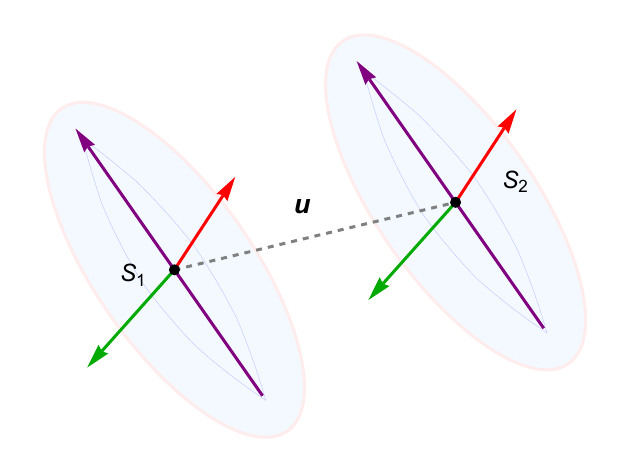}
\par\end{centering}
} \subfloat[\centering]{\includegraphics[width=0.5\columnwidth]{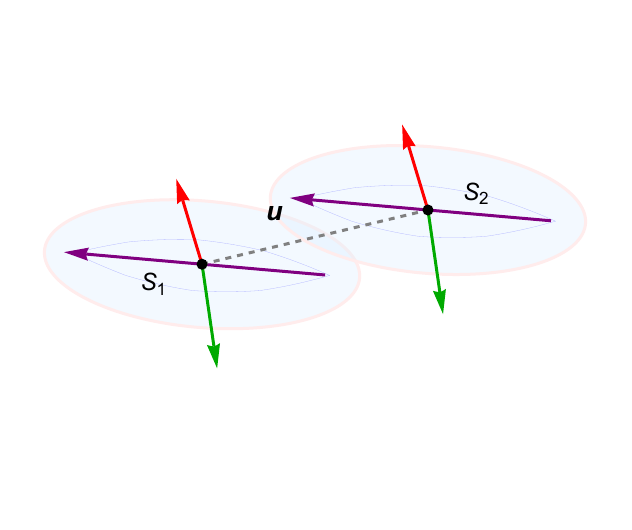}

}
\par\end{centering}
\caption{\label{fig:model2}Schematic illustration of two anisotropic Gaussian
point-spread functions centered at two fixed point-source positions.
The positions of the two sources and the \textit{\emph{relative}}
displacement direction are kept fixed in all panels, while only the
principal axes of the point-spread functions are physically rotated.
Panels $(a)$--$(d)$ correspond to increasing rotation angles, representing
different orientations of the anisotropic point-spread function with
respect to the displacement direction. As the relative orientation
of the two sources changes, the effective width of the point-spread
function along the source displacement varies, leading to different
spatial overlaps between the two displaced point-spread functions
and consequently different quantum-limited distance-estimation precisions.}
\end{figure}

It can be immediately seen that the rotation of the point-spread function
by $R$ is equivalent to rotating the relative orientation $\boldsymbol{u}$
between the two point sources by $R^{\mathsf{T}}$. Therefore, one
can optimize the resolution of the two point sources by rotating the
imaging system. It should be noted that such an approach works only
when the point-spread function is anisotropic, otherwise the matrix
representation of the response tensor $\boldsymbol{K}$ becomes proportional
to the identity matrix and the rotation $R$ cannot change $\boldsymbol{K}$.
Fig. \ref{fig:model2} illustrates the equivalence between rotating
the relative orientation $\boldsymbol{u}$ and rotating the point-spread
function $\psi(\boldsymbol{s})$ explicitly.

\subsection{Reflection symmetry structure of the point-spread function\label{subsec:Reflection-symmetry-structure}}

The above result can be further simplified when the point-spread function
has reflection symmetry with respect to a direction denoted by an
normalized vector $\boldsymbol{e}$.

The reflection that reverses the component of a vector $\boldsymbol{s}$
along $\boldsymbol{e}$ is
\begin{equation}
T_{\boldsymbol{e}}=I-2\boldsymbol{e}\boldsymbol{e}^{\intercal},\label{eq:ref}
\end{equation}
satisfying
\begin{equation}
T^{\intercal}_{\boldsymbol{e}}=T_{\boldsymbol{e}},\quad T^{\intercal}_{\boldsymbol{e}}T_{\boldsymbol{e}}=I,\;T_{\boldsymbol{e}}\boldsymbol{e}=-\boldsymbol{e}.\label{eq:t}
\end{equation}
Reflection symmetry of a point-spread function $\psi(\boldsymbol{s})$
about the axis $\boldsymbol{e}$ means
\begin{equation}
\psi(T_{\boldsymbol{e}}\boldsymbol{s})=\psi(\boldsymbol{s}).
\end{equation}

It can be verified that the gradient of the reflected point-spread
function $\psi(T_{\boldsymbol{e}}\boldsymbol{s})$ satisfies
\begin{equation}
\boldsymbol{\nabla}\psi(T_{\boldsymbol{e}}\boldsymbol{s})=T_{\boldsymbol{e}}\boldsymbol{\nabla}\psi(T_{\boldsymbol{e}}\boldsymbol{s}),
\end{equation}
therefore, the response tensor $\boldsymbol{K}$ associated with the
reflected point-spread function is
\begin{equation}
\begin{aligned}\int d^{3}\boldsymbol{s}\,[\boldsymbol{\nabla}\psi(T_{\boldsymbol{e}}\boldsymbol{s})][\boldsymbol{\nabla}\psi(T_{\boldsymbol{e}}\boldsymbol{s})]^{\mathsf{T}} & =T^{\intercal}_{\boldsymbol{e}}\boldsymbol{K}T_{\boldsymbol{e}}.\end{aligned}
\label{eq:K}
\end{equation}
Since the point-spread function $\psi(\boldsymbol{s})$ is invariant
under the reflection, the response tensor is unchanged, i.e.,
\begin{equation}
\boldsymbol{K}=T^{\intercal}_{\boldsymbol{e}}\boldsymbol{K}T_{\boldsymbol{e}}.
\end{equation}

This condition immediately leads to that a reflection symmetry axis
$\boldsymbol{e}$ of the point-spread function $\psi(\boldsymbol{s})$
is an eigenvector of $\boldsymbol{K}$. Indeed,
\begin{equation}
\boldsymbol{K}\boldsymbol{e}=T^{\intercal}_{\boldsymbol{e}}\boldsymbol{K}T_{\boldsymbol{e}}\boldsymbol{e}=-T_{\boldsymbol{e}}\boldsymbol{K}\boldsymbol{e},
\end{equation}
where the property of reflection operation $T_{\boldsymbol{e}}$ \eqref{eq:t}
is applied. This implies that $\boldsymbol{K}\boldsymbol{e}$ is an
eigenvector of $T_{\boldsymbol{e}}$ associated with the eigenvalue
$-1$. According to the definition of $T_{\boldsymbol{e}}$ \eqref{eq:ref},
$T_{\boldsymbol{e}}$ has only one non-degenerate eigenvalue $-1$
associated with the eigenvector $\boldsymbol{e}$, so $\boldsymbol{K}\boldsymbol{e}$
must be proportional to $\boldsymbol{e}$, i.e.,
\begin{equation}
\boldsymbol{K}\boldsymbol{e}=\lambda_{\boldsymbol{e}}\boldsymbol{e},
\end{equation}
where $\lambda_{\boldsymbol{e}}$ is a constant number which is actually
the eigenvalue of $\boldsymbol{K}$ associated with $\boldsymbol{e}$.
This verifies that the reflection axis $\boldsymbol{e}$ must be an
eigenvector of the response tensor $\boldsymbol{K}$.

Now, if the point-spread function $\psi(\boldsymbol{s})$ possesses
only one reflection symmetry axis, e.g., $\boldsymbol{e}_{1}$. In
this case, we can choose two additional orthonormal vectors, e.g.,
$\boldsymbol{e}_{2}$ and $\boldsymbol{e}_{3}$, spanning the two-dimensional
subspace orthogonal to $\boldsymbol{e}_{1}$. Based on the above results,
as the point-spread function $\psi(\boldsymbol{s})$ is symmetric
about $\boldsymbol{e}_{1}$, we have
\begin{equation}
\boldsymbol{K}\boldsymbol{e}_{1}=\lambda_{1}\boldsymbol{e}_{1},
\end{equation}
where $\lambda_{1}$ is the eigenvalue of $\boldsymbol{K}$ associated
with $\boldsymbol{e}_{1}$. In the orthonormal basis $\{\boldsymbol{e}_{1},\boldsymbol{e}_{2},\boldsymbol{e}_{3}\}$,
the matrix representation of the response tensor $\boldsymbol{K}$
has a block-diagonal representation,
\begin{equation}
\boldsymbol{K}=\left[\begin{array}{ccc}
\lambda_{1} & 0 & 0\\
0 & A & B\\
0 & B & C
\end{array}\right].
\end{equation}
In this case, one can immediately find the other two eigenvectors
of $\boldsymbol{K}$ in the subspace orthogonal to $\boldsymbol{e}_{1}$
and determine the optimal and worst orientations for the two optical
point sources among these three eigenvectors.

If the point-spread function $\psi(\boldsymbol{s})$ possesses a second
reflection axis $\boldsymbol{e}_{2}$, orthogonal to $\boldsymbol{e}_{1}$,
$\boldsymbol{e}_{2}$ is also an eigenvector of $\boldsymbol{K}$
then, associated with an eigenvalue denoted as $\lambda_{2}$. In
this case, the tensor $\boldsymbol{K}$ becomes completely diagonal
in the basis $\{\boldsymbol{e}_{1},\boldsymbol{e}_{2},\boldsymbol{e}_{3}\}$,
\begin{equation}
\boldsymbol{K}=\left[\begin{array}{ccc}
\lambda_{1} & 0 & 0\\
0 & \lambda_{2} & 0\\
0 & 0 & \lambda_{3}
\end{array}\right],
\end{equation}
where $\boldsymbol{e}_{3}$ is orthogonal to both $\boldsymbol{e}_{1}$
and $\boldsymbol{e}_{2}$ and $\lambda_{3}$ is its associated eigenvalue
of $\boldsymbol{K}$. It should be noted that while $\boldsymbol{e}_{3}$
is an eigenvector of $\boldsymbol{K}$ and thus $\boldsymbol{K}=T^{\intercal}_{3}\boldsymbol{K}T_{3}$
holds in this case, it does \emph{not} mean that $\boldsymbol{e}_{3}$
is necessarily a reflection symmetry axis of the point-spread function
$\psi(\boldsymbol{s})$. After all, the tensor $\boldsymbol{K}$ \eqref{eq:K}
involves integral of $\psi(\boldsymbol{s})$ over the whole three-dimensional
real space, so the symmetry of $\boldsymbol{K}$ does not imply the
same symmetry of $\psi(\boldsymbol{s})$.

The above result tells that the tensor $\boldsymbol{K}$ can generally
possess more symmetries than the point-spread function $\psi(\boldsymbol{s})$,
which provides facility for identifying the principal response directions
of displacement response. These symmetries turns out to determine
the eigenvectors of $\boldsymbol{K}$ and hence the optimal and worst
directions for the orientation of the two point sources. The ratio
of the largest and smallest eigenvalues then characterize the extent
that the resolution can be increased by properly rotating the imaging
system, the as discussed in Sec. \ref{subsec:Orientational-enhancement}.

\section{Example: two Gaussian point-spread functions}

In this section, we consider a general three-dimensional Gaussian
point-spread function as an example, and examine how the spatial structure
of the point-spread function determines the quantum-limited estimation
precision of the distance between two incoherent point sources.

Suppose the point-spread-function of the $i$-th source centered at
$\boldsymbol{s}_{i}$ is
\begin{equation}
\psi_{i}(\boldsymbol{s})=\psi(\boldsymbol{s}-\boldsymbol{s}_{i}),\quad i=1,2,
\end{equation}
where $\psi(\boldsymbol{s})$ is the \emph{amplitude} of a real three-dimensional
Gaussian distribution,

\begin{equation}
\psi(\boldsymbol{s})=\frac{\exp\left[-\frac{1}{4}\boldsymbol{s}^{\intercal}\boldsymbol{\Sigma}^{-1}\boldsymbol{s}\right]}{(2\pi)^{3/4}\left|\boldsymbol{\Sigma}\right|^{1/4}}.\label{eq:gaussian}
\end{equation}
Denote the width of the Gaussian distribution in the $x,y,z$ directions
as $w_{x},w_{y},w_{z}$ respectively. The covariance matrix $\boldsymbol{\Sigma}$
can then be decomposed as
\begin{equation}
\boldsymbol{\Sigma}=\boldsymbol{WCW}^{\mathsf{T}},
\end{equation}
where $\boldsymbol{W}=\text{diag}(w_{x},w_{y},w_{z})$ and
\begin{equation}
\boldsymbol{C}=\left(\begin{array}{ccc}
1 & \beta_{xy} & \beta_{xz}\\
\beta_{xy} & 1 & \beta_{yz}\\
\beta_{xz} & \beta_{yz} & 1
\end{array}\right)>0,
\end{equation}
specifies the correlations of the Gaussian distribution between the
$x,y,z$ directions in the laboratory coordinate system. The condition
$\boldsymbol{C}>0$ ensures that $\boldsymbol{\Sigma}$ is positive
definite.

A key result of superresolution for Gaussian point-spread functions
is that, as shown in Appendix \ref{subsec:3dqf}, the matrix representation
of the response tensor $\boldsymbol{K}$ is directly determined by
the covariance matrix $\boldsymbol{\Sigma}$,
\begin{equation}
\boldsymbol{K}=\frac{1}{4}\boldsymbol{\Sigma}^{-1}.\label{eq:ksigma}
\end{equation}
This immediately gives the distance precision \eqref{eq:hr} in the
sub-Rayleigh limit as
\begin{equation}
\hr[\text{}]=\frac{\nt\left(1-\epsilon^{2}\right)}{4\boldsymbol{u}^{\intercal}\boldsymbol{\Sigma}\boldsymbol{u}}.\label{eq:ghr}
\end{equation}

Suppose $\boldsymbol{n}_{1},\boldsymbol{n}_{2},\boldsymbol{n}_{3}$
are the normalized vectors of the three orthogonal principal axes
of the Gaussian distribution, satisfying
\begin{equation}
\boldsymbol{\Sigma}\boldsymbol{n}_{i}=t_{i}\boldsymbol{n}_{i},\;i=1,2,3.
\end{equation}
If the relative displacement $\boldsymbol{u}$ is along one of the
principal axis of the Gaussian distribution, e.g., $\boldsymbol{u}\propto\boldsymbol{n}_{i},$the
precision $\hr[\text{}]$ can be simplified as
\begin{equation}
\hr[\text{}]=\frac{\nt\left(1-\epsilon^{2}\right)}{4t_{i}}.\label{eq:hti}
\end{equation}
Moreover, the Gaussian point-spread function is symmetric about the
principal axes $\boldsymbol{n}_{1},\boldsymbol{n}_{2},\boldsymbol{n}_{3}$,
therefore, by the results of Sec. \ref{subsec:Reflection-symmetry-structure},
the response tensor $\boldsymbol{K}$ can be diagonalized in the basis
$\{\boldsymbol{n}_{1},\boldsymbol{n}_{2},\boldsymbol{n}_{3}\}$. This
is in accordance with Eq. \eqref{eq:ksigma} which implies that the
response tensor $\boldsymbol{K}$ shares the same eigenvectors with
the covariance matrix $\boldsymbol{\Sigma}$.

Now, let $t_{\min}$ and $t_{\max}$ denote the smallest and largest
eigenvalues of the covariance matrix $\Sigma$, with the principal
directions $\boldsymbol{n}_{\min}$ and $\boldsymbol{n}_{\max}$.
The optimal relative orientation between the two point sources is
obtained when
\begin{equation}
\boldsymbol{u}_{\mathrm{opt}}=\boldsymbol{n}_{\min},
\end{equation}
whereas the worst direction is obtained when
\begin{equation}
\boldsymbol{u}_{\mathrm{wor}}=\boldsymbol{n}_{\max}.
\end{equation}
The highest and lowest precisions are then given by
\begin{equation}
\dlm{\hr[({\rm opt}/{\rm wor})]}=\frac{\nt\left(1-\epsilon^{2}\right)}{4t_{\min/\max}}.\label{eq:hwg}
\end{equation}
Thus, the spectrum of the covariance matrix determines the range of
distance precisions accessible through different relative orientations
of the two sources with Gaussian point-spread functions. The maximum
ratio that the resolution can be increased by rotating the point sources
or equivalently the imaging system is therefore
\begin{equation}
\xi=\frac{t_{\max}}{t_{\min}},
\end{equation}
which means that the more anisotropic the Gaussian distribution is,
the larger enhancement ratio that the resolution can increased.

These results have an intuitive geometric interpretation: when the
point-spread function has a smaller spatial spread, the distance information
is enhanced along the minor axis of the Gaussian point-spread function,
as it would produce a larger change of the optical field under displacement.
\begin{figure}[!b]
\begin{centering}
\subfloat[\centering\label{fig:ap1long}]{\includegraphics[width=0.52\columnwidth]{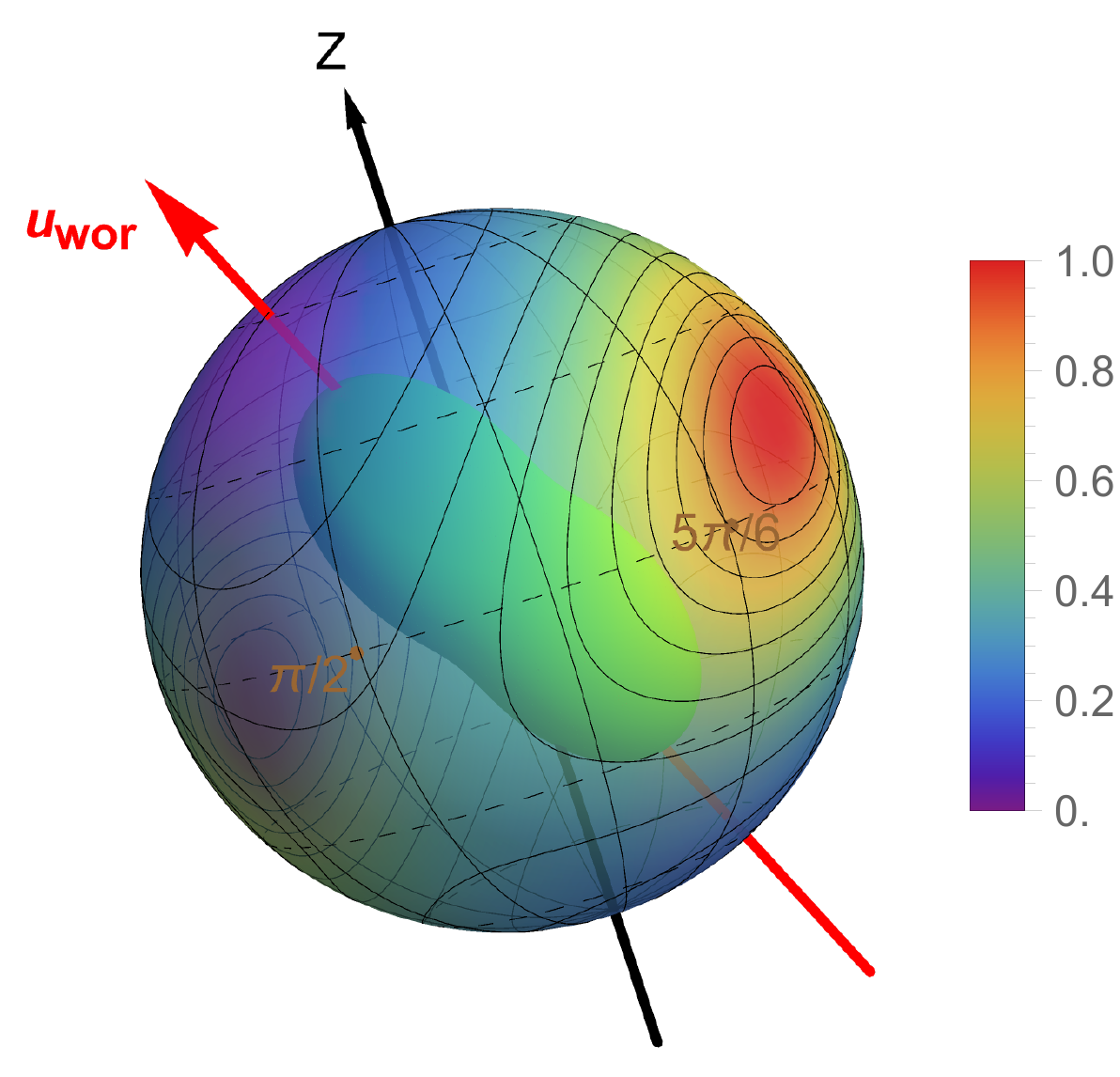}}\subfloat[\centering\label{fig:ap2short}]{\includegraphics[width=0.52\columnwidth]{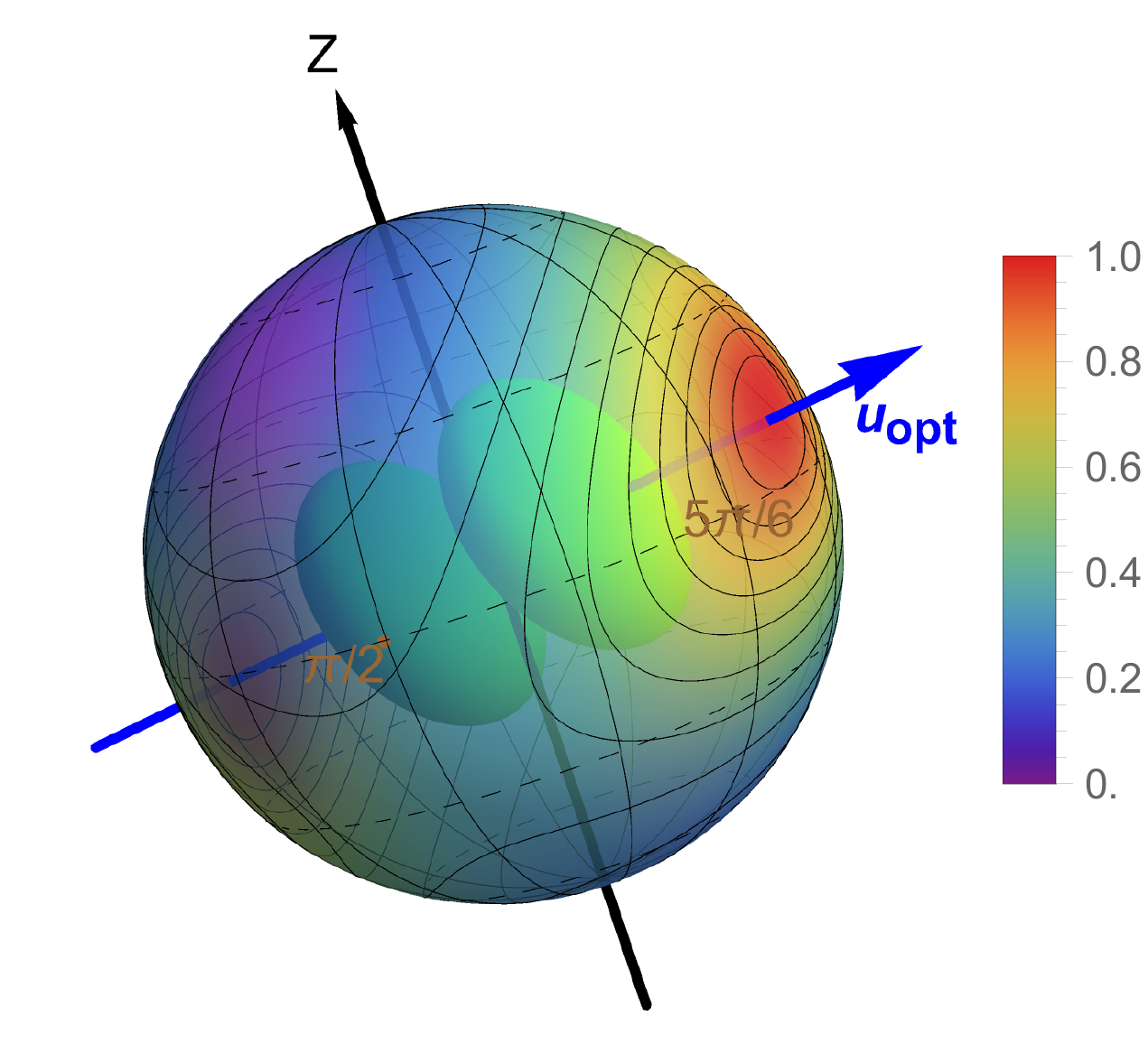}

}
\par\end{centering}
\caption{\label{fig:ap} Angular dependence of the quantum-limited precision
$\protect\hr$ for estimating the distance between two balanced incoherent
point sources in the limit $r\rightarrow0$, shown as a function of
the azimuthal angle $\alpha$ and the polar angle $\phi$. The precision
is normalized by $\protect\nt/(4w^{2}_{x})$. The point-spread function
is Gaussian with $w_{y}/w_{x}=1.5$, $w_{z}/w_{x}=1.8$ and correlation
coefficients $\beta_{xy}=0.25$, $\beta_{yz}=0.2$, and $\beta_{xz}=0.3$.
The angular dependence reflects the anisotropic spatial structure
of the Gaussian point-spread function: the precision is maximized
when the source displacement orientation is aligned with the minor
principal axis of the Gaussian covariance ellipsoid as shown by Fig.
\ref{fig:ap2short} and minimized along its major principal axis as
shown by Fig. \ref{fig:ap1long}.}
\end{figure}

When the Gaussian distribution is isotropic, the covariance matrix
becomes proportional to the identity matrix,
\begin{equation}
\boldsymbol{\Sigma}=w^{2}\boldsymbol{I},
\end{equation}
where $w$ is the width of the Gaussian distribution along any direction.
In this case, the overlap between the two point-spread functions remains
invariant with their \textit{relative} orientation, and the estimation
precision becomes
\begin{equation}
\hr[{\rm cir}]=\frac{\nt\left(1-\epsilon^{2}\right)}{4w^{2}},
\end{equation}
which is independent of the relative orientation. Therefore, no orientation-dependent
precision gain can be acquired for this case.

\begin{figure}[!t]
\centering{}\includegraphics[scale=0.55]{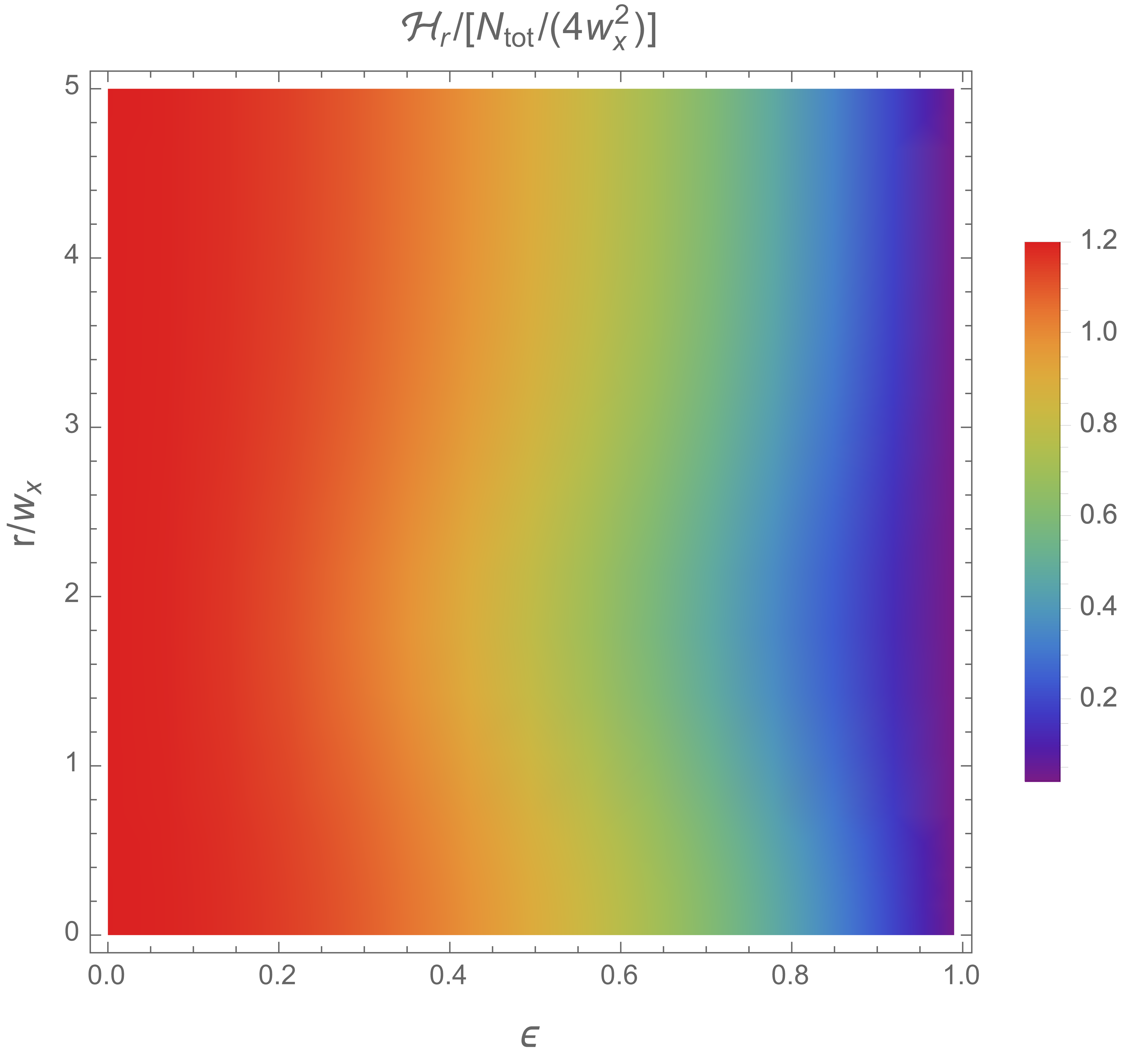}\caption{\label{fig:re} Ratio $\protect\hr/[\protect\nt/(4w^{2}_{x})]$ as
a function of the distance $r$ and the intensity imbalance parameter
$\epsilon$, evaluated at the optimal orientation $(\alpha,\phi)\approx(2.95,1.40)$.
The point-spread function is Gaussian with $w_{y}/w_{x}=1.5,w_{z}/w_{x}=1.8$,
and correlation coefficients $\beta_{xy}=0.25$, $\beta_{yz}=0.2$,
and $\beta_{xz}=0.3$. It can be observed that for a given distance
$r$, the precision decreases with increasing intensity imbalance
and attains its maximum at $\epsilon=0$, corresponding to equal photon
numbers from the two sources. In contrast, for a fixed imbalance parameter
$\epsilon$, the precision exhibits only weak dependence on the distance
$r$ in the sub-Rayleigh regime shown, reflecting the absence of Rayleigh's
curse at the quantum limit.}
\end{figure}

The estimation precision for the distance between two balanced incoherent
point sources with a Gaussian point-spread function is numerically
illustrated in the limit $r\rightarrow0$ in Fig. \ref{fig:ap}, which
clearly shows the dependence of the precision on the \textit{relative}
displacement direction between the two sources and that the optimal
and worst directions coincide with the principal axes of the Gaussian
distribution, in accordance with the relation $\hr[\text{}]\propto1/t_{i}$
in Eq. \eqref{eq:hti}. The figure can therefore be read as a reorientation
of the two point sources relative to a fixed displacement direction:
In a fixed laboratory frame, the source positions determine $\boldsymbol{u}$,
while the response tensor is changed as $\boldsymbol{K}_{R}=R\boldsymbol{K}R^{\mathsf{T}}$
by a rotation $R$ of the imaging system. The optimal configuration
is reached when the minor principal axis of the rotated Gaussian point-spread
function is along the source displacement $\boldsymbol{u}$. The analysis
is further extended for two unbalanced point sources with finite distance
in Fig. \ref{fig:re}. In the small-$r$ regime, the behavior of the
precision follows $\hr\propto1-\epsilon^{2}$ according to Eq. \eqref{eq:ghr},
indicating that intensity imbalance reduces the available information
of the distance. In the strongly imbalanced regime $\epsilon\rightarrow1$,
the precision is suppressed and finally vanishes as one of point sources
dominates.

\section{Conclusion}

In this work, we have established a general geometric framework for
quantum-limit distance estimation between two incoherent point sources
in a three-dimensional imaging system. We have shown that, in the
sub-Rayleigh regime, the information of the source distance remains
finite and is determined by the second-order displacement-response
tensor of the point-spread function. Beyond demonstrating the absence
of Rayleigh's curse at the quantum limit, our results reveal that
three-dimensional quantum superresolution possesses an intrinsic geometric
structure: the eigenvectors of the displacement-response tensor define
the principal directions of source displacement, while the corresponding
eigenvalues determine the achievable distance-estimation precision.

This geometric viewpoint provides a direct connection between the
spatial structure of the imaging system and the ultimate estimation
performance. For anisotropic point-spread functions, the optimal and
least favorable precisions can be accessed by physically reorienting
the imaging system relative to the source displacement. We further
show that reflection symmetries of the point-spread function can identify
the principal displacement-response directions without explicitly
evaluating the full tensor. For Gaussian point-spread functions, this
framework is reduced to a simple linear relation between the displacement-response
tensor and the inverse spatial covariance matrix, connecting the attainable
precision directly to the spatial geometry of the point-spread function.

These results provide a general geometric characterization of three-dimensional
quantum superresolution and demonstrate how the spatial structure
and orientation of the point-spread function determines the ultimate
precision of distance estimation, revealing the potential of point-spread
function engineering for improving quantum superresolution.

\section*{ACKNOWLEDGMENTS}

This work is supported by the National Natural Science Foundation
of China (Grant No. 12075323), the Natural Science Foundation of Guangdong
Province of China (Grant No. 2025A1515011440) and the Innovation Program
for Quantum Science and Technology (Grant no. 2021ZD0300702).

\bibliographystyle{apsrev4-2}
\bibliography{superresolution}

\appendix
\onecolumngrid

\section{Precision of quantum superresolution\label{sec:Precision-of-quantum}}

\subsection{\label{subsec:com3}Quantum precision limit for Cartesian components
of the displacement in three-dimensional system}

Building upon the pioneering quantum superresolution framework for
one-dimensional imaging systems developed by Tsang\textit{ et al.}
\citep{Tsang2016}, we extend the superresolution theory to two incoherent
optical point sources with arbitrary intensities. The quantum state
of a single detected photon can be described by the density operator
\begin{equation}
\begin{aligned}\hro= & \frac{1-\epsilon}{2}\ksi[1]\bsi[1]+\frac{1+\epsilon}{2}\ksi[2]\bsi[2]\end{aligned}
,\label{eq:ro}
\end{equation}
where $\ksi[i]=\nei[x]{X_{i}}\nei[y]{Y_{i}}\nei[z]{Z_{i}}\ksi,i=1,2$,
and $\epsilon$ is the normalized intensity difference between the
two incoherent sources. Let $\boldsymbol{g}=(g_{1},\ldots,g_{k})$
denote the spatial parameters to be estimated. Any physically admissible
measurement is described by a POVM and produces a classical Fisher
information matrix. The covariance matrix of any locally unbiased
estimator obeys
\begin{equation}
{\rm Cov}[\hat{\boldsymbol{g}}]\geq\nt[-1]\mathcal{Q}^{-1}[\boldsymbol{g}],
\end{equation}
where $\mathcal{Q}$ is the quantum Fisher information matrix, defined
as
\begin{equation}
\mathcal{Q}_{ij}(\hro)=\frac{1}{2}{\rm Tr}[\hro\{\hl{g_{i}},\hl{g_{j}}\}],\forall g_{i},g_{j}\in\boldsymbol{g},\label{eq:fisher0-1}
\end{equation}
where $\{\cdot,\cdot\}$ denotes the anticommutator and $\hl{g_{i}}$
is the symmetric logarithmic derivative (SLD) of the density operator
$\hro$ with respect to the parameter $g_{i}$, 
\begin{equation}
\frac{1}{2}(\hl{g_{i}}\hro+\hro\hl{g_{i}})=\partial_{g_{i}}\hro.\label{eq:sld}
\end{equation}

The single-photon state is supported on the subspace
\begin{equation}
\mathcal{S}=\text{span}\{\ksi[1],\ksi[2]\},
\end{equation}
and therefore has rank at most two, independently of the spatial dimension.
For two distinct source modes with both sources present, its rank
is two. Defining the overlap
\begin{equation}
\delta\equiv\left\langle \psi_{1}\mid\psi_{2}\right\rangle \neq0.
\end{equation}
an orthonormal basis of $\mathcal{S}$ can be chosen as
\begin{align}
|\varphi_{1,2}\rangle & =\frac{1}{\sqrt{\mathscr{Q}_{1,2}}}\times\{\ksi[1]\mp\frac{\left(\sqrt{\epsilon^{2}+\delta^{2}\left(1-\epsilon^{2}\right)}\mp\epsilon\right)}{\delta(1-\epsilon)}\ksi[2]\},\label{eq:es}
\end{align}
where $\mathscr{Q}_{1,2}$ are the normalization constants,
\begin{align}
\mathscr{Q}_{1,2} & =\left(1-\delta^{2}\right)\left(1+\frac{\delta^{2}\left(1-\delta^{2}\right)(1-\epsilon)^{2}}{\left(\sqrt{\delta^{2}+\epsilon^{2}-\delta^{2}\epsilon^{2}}\pm\delta^{2}(1-\epsilon)\pm\epsilon\right)^{2}}\right).
\end{align}
The corresponding eigenvalues are
\begin{equation}
\varphi_{1,2}=\frac{1}{2}\left(1\mp\sqrt{\epsilon^{2}+\delta^{2}\left(1-\epsilon^{2}\right)}\right).\label{eq:ev}
\end{equation}
Using the spectral decomposition $\hro=\sum_{k}\varphi_{k}|\varphi_{k}\rangle\langle\varphi_{k}|$,
an SLD $\hl{g_{i}}$ can be expressed as
\begin{equation}
\hl{g_{i}}=\sum_{\varphi_{k}+\varphi_{h}\neq0}\frac{2\langle\varphi_{k}|\partial_{g_{i}}\hro|\varphi_{h}\rangle}{\varphi_{k}+\varphi_{h}}\ket{\varphi_{k}}\bra{\varphi_{h}}.
\end{equation}

The quantum Fisher information matrix is more conveniently calculated
within the support of $\hro$,
\begin{equation}
\mathcal{Q}_{ij}(\hat{\rho})=\sum_{\varphi_{k}\neq0}\frac{4\langle\varphi_{k}|\partial_{g_{i}}\hro\partial_{g_{j}}\hro|\varphi_{k}\rangle}{\varphi_{k}}+\sum_{\varphi_{k},\varphi_{h}\neq0}2\left(\frac{1}{\varphi_{k}+\varphi_{h}}-\frac{1}{\varphi_{k}}-\frac{1}{\varphi_{h}}\right)\langle\varphi_{h}|\partial_{g_{i}}\hro|\varphi_{k}\rangle\langle\varphi_{k}|\partial_{g_{j}}\hro|\varphi_{h}\rangle,\,\varphi_{k},\varphi_{h}\in\{\varphi_{1},\varphi_{2}\},\label{eq:fisher exp}
\end{equation}
where the summations run over the two nonzero eigenvalues $\varphi_{1}$
and $\varphi_{2}$ of $\hro$.

In multiparameter quantum estimation, the symmetric-logarithmic-derivative
quantum Cram\'er-Rao bound is not generally jointly attainable, because
measurements optimal for different parameters may be incompatible.
To quantify the estimation performance for a specified combination
of parameters, we introduce a positive-semidefinite weight matrix
$\boldsymbol{W}$ and define the scalar covariance cost ${\rm Tr}(\boldsymbol{W}{\rm Cov}[\hat{\boldsymbol{g}}])$,
the SLD quantum Cram\'er-Rao bound then gives

\begin{equation}
{\rm Tr}(\boldsymbol{W}{\rm Cov}[\hat{\boldsymbol{g}}])\geq\nt[-1]{\rm Tr}(\boldsymbol{W}\mathcal{Q}^{-1}[\boldsymbol{g}]).\label{eq:scalar-1}
\end{equation}

Although this scalar bound is not attainable for a general multiparameter
model, it coincides with the asymptotically attainable Holevo bound
when the weak commutativity condition

\begin{equation}
{\rm Tr}(\hro[\hl{g_{i}},\hl{g_{j}}])={\rm ImTr}(\hro\hl{g_{i}}\hl{g_{j}})=0,\label{eq:compatibility-1}
\end{equation}
holds for every pair of parameters $g_{i}$, $g_{j}$ assigned nonzero
weight \citep{Ragy2016}. For the real point-spread function considered
here, $\hro$ and its parameter derivatives are real symmetric in
the position basis, and the corresponding SLDs may also be chosen
real symmetric. The above weak commutativity condition is therefore
satisfied. Consequently, the weighted SLD quantum Cram\'er-Rao bound
provides an asymptotically attainable precision limit for the present
superresolution model \citep{Ragy2016,Rehacek2017}.

In a three-dimensional imaging system, the centroid and displacement
are described by
\begin{equation}
\overline{\boldsymbol{s}}=(\bar{X},\bar{Y},\zb),\,\Delta\boldsymbol{s}=(d_{x},d_{y},d_{z})^{\top}.
\end{equation}
Accordingly, the quantum state $\hro$ depends on the parameter vector
\begin{equation}
\boldsymbol{g}=(\xb,\yb,\zb,d_{x},d_{y},d_{z})^{\top}.
\end{equation}
The quantum Fisher information matrix with respect to unknown parameters
$\boldsymbol{g}$ can be derived by plugging the eigenvalues \eqref{eq:ev}
and eigenstates \eqref{eq:es} of the density matrix $\hro$ into
Eq. \eqref{eq:fisher exp}, and is given by
\begin{equation}
Q=\left(\begin{array}{cccccc}
4\kappa_{x}-4\left(1-\epsilon^{2}\right)\gamma^{2}_{x} & \,4\et[xy]-4\left(1-\epsilon^{2}\right)\gamma_{x}\gamma_{y} & 4\et[xz]-4\left(1-\epsilon^{2}\right)\gamma_{x}\gamma_{z} & 2\epsilon\kappa_{x} & 2\epsilon\et[xy] & 2\epsilon\et[xz]\\
4\et[xy]-4\left(1-\epsilon^{2}\right)\gamma_{x}\gamma_{y} & 4\kappa_{y}-4\left(1-\epsilon^{2}\right)\gamma^{2}_{y} & 4\et[yz]-4\left(1-\epsilon^{2}\right)\gamma_{y}\gamma_{z} & 2\epsilon\et[xy] & 2\epsilon\kappa_{y} & 2\epsilon\et[yz]\\
4\et[xz]-4\left(1-\epsilon^{2}\right)\gamma_{x}\gamma_{z} & 4\et[yz]-4\left(1-\epsilon^{2}\right)\gamma_{y}\gamma_{z} & 4\kappa_{z}-4\left(1-\epsilon^{2}\right)\gamma^{2}_{z} & 2\epsilon\et[xz] & 2\epsilon\et[yz] & 2\epsilon\kappa_{z}\\
2\epsilon\kappa_{x} & 2\epsilon\et[xy] & 2\epsilon\et[xz] & \kappa_{x} & \et[xy] & \et[xz]\\
2\epsilon\et[xy] & 2\epsilon\kappa_{y} & 2\epsilon\et[yz] & \et[xy] & \kappa_{y} & \et[yz]\\
2\epsilon\et[xz] & 2\epsilon\et[yz] & 2\epsilon\kappa_{z} & \et[xz] & \et[yz] & \kappa_{z}
\end{array}\right),\label{eq:fimc}
\end{equation}
where
\begin{align}
\kappa_{i}= & \avg[i][][][2]=\ensuremath{\int d^{3}\boldsymbol{s}\,\partial_{i}\psi^{\ast}\,\partial_{i}\psi},\;i\in\{x,y,z\},\nonumber \\
\et[ij]= & \av[{\hp[i]\hp[j]}]=\ensuremath{\int d^{3}\boldsymbol{s}\,\partial_{i}\psi^{\ast}\,\partial_{j}\psi},\;i,j\in\{x,y,z\}\:i\neq j,\\
\delta= & \av[{\nei[x]{r{\rm cos}\alpha\sin\phi}\nei[y]{r{\rm sin}\alpha\sin\phi}\nei[z]{r\cos\phi}}]=\av[{\cos(r\hp[r])}],\\
\gamma_{x}= & i\av[{\hp[x]\nei[x]{r{\rm cos}\alpha\sin\phi}\nei[y]{r{\rm sin}\alpha\sin\phi}\nei[z]{r\cos\phi}}]=\av[{\hp[x]\sin(r\hp[r])}],\nonumber \\
\gamma_{y}= & i\av[{\hp[y]\nei[x]{r{\rm cos}\alpha\sin\phi}\nei[y]{r{\rm sin}\alpha\sin\phi}\nei[z]{r\cos\phi}}]=\av[{\hp[y]\sin(r\hp[r])}],\nonumber \\
\gamma_{z}= & i\av[{\hp[z]\nei[x]{r{\rm cos}\alpha\sin\phi}\nei[y]{r{\rm sin}\alpha\sin\phi}\nei[z]{r\cos\phi}}]=\av[{\hp[z]\sin(r\hp[r])}],\nonumber 
\end{align}
where $\hp[r]=\boldsymbol{u}^{\intercal}\hat{\boldsymbol{P}}=\hp[x]\cos\alpha\sin\phi+\hp[y]\sin\alpha\sin\phi+\hp[z]\cos\phi$
is the momentum operator along the \textit{relative} displacement
direction between the two sources. The quantum-limited precisions
for estimating the displacement components $d_{x}$, $d_{y}$ and
$d_{z}$ are given by the inverse QFI matrix,
\begin{equation}
\mathcal{H}_{d_{x}}=\frac{\nt\left(1-\epsilon^{2}\right)\mathscr{M}}{\kappa_{y}\kappa_{z}-\et[yz][2]},
\end{equation}
\begin{equation}
\mathcal{H}_{d_{y}}=\frac{\nt\left(1-\epsilon^{2}\right)\mathscr{M}}{\kappa_{x}\kappa_{z}-\et[xz][2]},
\end{equation}
\begin{equation}
\mathcal{H}_{d_{z}}=\frac{\nt\left(1-\epsilon^{2}\right)\mathscr{M}}{\kappa_{x}\kappa_{y}-\et[xy][2]}.
\end{equation}
A more compact structure emerges by recognizing that the component-wise
expressions for $\mathcal{H}_{d_{x}}$, $\mathcal{H}_{d_{y}}$, and
$\mathcal{H}_{d_{z}}$ can be unified through the symmetric second-moment
tensor $\boldsymbol{K}$
\begin{align}
\boldsymbol{K}= & \ensuremath{\int d^{3}\boldsymbol{s}\,[\boldsymbol{\nabla}\psi(\boldsymbol{s})][\boldsymbol{\nabla}\psi(\boldsymbol{s})]^{\mathsf{T}}=\left(\begin{array}{ccc}
\kappa_{x} & \et[xy] & \et[xz]\\
\et[xy] & \kappa_{y} & \et[yz]\\
\et[xz] & \et[yz] & \kappa_{z}
\end{array}\right),}
\end{align}
which is positive definite and fully characterizes the spatial response
of the PSF.

Explicitly,
\begin{equation}
\mathscr{M}=\det\boldsymbol{K}=\kappa_{x}\kappa_{y}\kappa_{z}+2\eta_{xy}\eta_{xz}\eta_{yz}-\kappa_{x}\eta^{2}_{yz}-\kappa_{y}\eta^{2}_{xz}-\kappa_{z}\eta^{2}_{xy},
\end{equation}
For a three-dimensional symmetric tensor $\boldsymbol{K}$, its inverse
can be expressed as
\begin{equation}
\boldsymbol{K}^{-1}=\frac{{\rm adj}\boldsymbol{K}}{\det\boldsymbol{K}},
\end{equation}
where ${\rm adj}\boldsymbol{K}$ denotes the adjugate matrix representation
of $\boldsymbol{K}$, the quantum-limited precisions for the displacement
components can then be rewritten as
\begin{equation}
\mathcal{H}_{d_{i}}=\nt(1-\epsilon^{2})/(\boldsymbol{K}^{-1})_{ii},\,i=x,y,z.
\end{equation}
This representation makes the structure transparent: $\boldsymbol{K}$
encodes the intrinsic second-order spatial properties of the PSF,
while its inverse governs the attainable precision along each Cartesian
direction. Its positive definiteness ensures finite, nonvanishing
precision for all components even in the limit of vanishing distance.

\subsection{\label{subsec:3dqf}Quantum estimation precision of source distance
in three dimensions}

It is therefore natural to parameterize the displacement vector in
spherical coordinates by its magnitude and direction. In this representation,
the density operator $\hro$ depends on the parameter set 
\begin{equation}
\boldsymbol{g}=(\xb,\yb,\zb,r,\alpha,\phi)^{\top}.
\end{equation}
The corresponding quantum Fisher information matrix can be obtained
from Eq. \eqref{eq:fisher exp} using the spectral decomposition of
$\hro$
\begin{equation}
Q=\left(\begin{array}{cccccc}
4\kappa_{x}-4\left(1-\epsilon^{2}\right)\gamma^{2}_{x} & \,4\et[xy]-4\left(1-\epsilon^{2}\right)\gamma_{x}\gamma_{y} & 4\et[xz]-4\left(1-\epsilon^{2}\right)\gamma_{x}\gamma_{z} & 2\epsilon\kp[xr] & 2r\epsilon\kp[x\alpha] & 2r\epsilon\kp[x\phi]\\
4\et[xy]-4\left(1-\epsilon^{2}\right)\gamma_{x}\gamma_{y} & 4\kappa_{y}-4\left(1-\epsilon^{2}\right)\gamma^{2}_{y} & 4\et[yz]-4\left(1-\epsilon^{2}\right)\gamma_{y}\gamma_{z} & 2\epsilon\kp[yr] & 2r\epsilon\kp[y\alpha] & 2r\epsilon\kp[y\phi]\\
4\et[xz]-4\left(1-\epsilon^{2}\right)\gamma_{x}\gamma_{z} & 4\et[yz]-4\left(1-\epsilon^{2}\right)\gamma_{y}\gamma_{z} & 4\kappa_{z}-4\left(1-\epsilon^{2}\right)\gamma^{2}_{z} & 2\epsilon\kp[zr] & 2r\epsilon\kp[z\alpha] & 2r\epsilon\kp[z\phi]\\
2\epsilon\kp[xr] & 2r\epsilon\kp[x\alpha] & 2r\epsilon\kp[x\phi] & \kappa_{r} & r\kp[r\alpha] & r\kp[r\phi]\\
2\epsilon\kp[yr] & 2r\epsilon\kp[y\alpha] & 2r\epsilon\kp[y\phi] & r\kp[r\alpha] & r^{2}\kappa_{\alpha} & r^{2}\kp[\alpha\phi]\\
2\epsilon\kp[zr] & 2r\epsilon\kp[z\alpha] & 2r\epsilon\kp[z\phi] & r\kp[r\phi] & r^{2}\kp[\alpha\phi] & r^{2}\kappa_{\phi}
\end{array}\right),\label{eq:qfi3}
\end{equation}
where
\begin{equation}
\begin{aligned}\kappa_{r}= & \boldsymbol{u}^{\intercal}\boldsymbol{K}\boldsymbol{u},\\
= & \kappa_{x}\cos^{2}\alpha\sin^{2}\phi+\kappa_{y}\sin^{2}\alpha\sin^{2}\phi+\kappa_{z}\cos^{2}\phi+\eta_{xy}\sin2\alpha\sin^{2}\phi+\eta_{xz}\cos\alpha\sin2\phi+\eta_{yz}\sin\alpha\sin2\phi,\\
\kappa_{\alpha}= & \sin^{2}\phi\boldsymbol{v^{\intercal}_{\alpha}}\boldsymbol{K}\boldsymbol{v_{\alpha}}\\
= & \sin^{2}\phi\left(\kappa_{x}\sin^{2}\alpha-\eta_{xy}\sin2\alpha+\kappa_{y}\cos^{2}\alpha\right),\\
\kappa_{\phi}= & \boldsymbol{v^{\intercal}_{\phi}}\boldsymbol{K}\boldsymbol{v_{\phi}},\\
= & \kappa_{x}\cos^{2}\alpha\cos^{2}\phi+\kappa_{y}\sin^{2}\alpha\cos^{2}\phi+\kappa_{z}\sin^{2}\phi+\eta_{xy}\sin2\alpha\cos^{2}\phi-\eta_{xz}\cos\alpha\sin2\phi-\eta_{yz}\sin\alpha\sin2\phi,
\end{aligned}
\end{equation}
and the correlation quantities
\begin{equation}
\begin{aligned}\kp[r\alpha]= & \sin\phi\boldsymbol{v^{\intercal}_{\alpha}}\boldsymbol{K}\boldsymbol{u},\\
= & \frac{1}{2}\kappa_{y}\sin2\alpha\sin^{2}\phi+\frac{1}{2}\eta_{yz}\cos\alpha\sin2\phi-\frac{1}{2}\kappa_{x}\sin2\alpha\sin^{2}\phi+\eta_{xy}\cos2\alpha\sin^{2}\phi-\frac{1}{2}\eta_{xz}\sin\alpha\sin2\phi,\\
\kappa_{r\phi}= & \boldsymbol{v^{\intercal}_{\phi}}\boldsymbol{K}\boldsymbol{u},\\
= & \frac{1}{2}\kappa_{x}\cos^{2}\alpha\sin2\phi+\frac{1}{2}\kappa_{y}\sin^{2}\alpha\sin2\phi-\frac{1}{2}\kappa_{z}\sin2\phi+\frac{1}{2}\eta_{xy}\sin2\alpha\sin2\phi+\eta_{xz}\cos\alpha\cos2\phi+\eta_{yz}\sin\alpha\cos2\phi,\\
\kappa_{\alpha\phi}= & \sin\phi\boldsymbol{v^{\intercal}_{\alpha}}\boldsymbol{K}\boldsymbol{v_{\phi}},\\
= & \frac{1}{2}\eta_{xy}\cos2\alpha\sin2\phi-\frac{1}{4}\kappa_{x}\sin2\alpha\sin2\phi+\eta_{xz}\sin\alpha\sin^{2}\phi+\frac{1}{4}\kappa_{y}\sin2\alpha\sin2\phi-\eta_{yz}\cos\alpha\sin^{2}\phi.
\end{aligned}
\end{equation}
The unit vectors along the $x-$, $y-$ and $z-$axes are denoted
by
\begin{equation}
\boldsymbol{e}_{x}=(1,0,0)^{\intercal},\boldsymbol{e}_{y}=(0,1,0)^{\intercal},\boldsymbol{e}_{z}=(0,0,1)^{\intercal},
\end{equation}
then the quantities in Eq. \eqref{eq:qfi3} can be written as
\begin{equation}
\begin{aligned}\kp[xr]= & \boldsymbol{e}^{\intercal}_{x}\boldsymbol{K}\boldsymbol{u}=\et[xy]\sin\alpha\sin\phi+\kappa_{x}\cos\alpha\sin\phi+\et[xz]\cos\phi,\\
\kp[x\alpha]= & \sin\phi\boldsymbol{e}^{\intercal}_{x}\boldsymbol{K}\boldsymbol{v_{\alpha}}=\sin\phi\left(\et[xy]\cos\alpha-\kappa_{x}\sin\alpha\right),\\
\kp[x\phi]= & \boldsymbol{e}^{\intercal}_{x}\boldsymbol{K}\boldsymbol{v_{\phi}}=\et[xy]\sin\alpha\cos\phi+\kappa_{x}\cos\alpha\cos\phi-\et[xz]\sin\phi,\\
\kp[yr]= & \boldsymbol{e}^{\intercal}_{y}\boldsymbol{K}\boldsymbol{u}=\et[xy]\cos\alpha\sin\phi+\et[yz]\cos\phi+\kappa_{y}\sin\alpha\sin\phi,\\
\kp[y\alpha]= & \sin\phi\boldsymbol{e}^{\intercal}_{y}\boldsymbol{K}\boldsymbol{v_{\alpha}}=\sin\phi\left(\kappa_{y}\cos\alpha-\et[xy]\sin\alpha\right),\\
\kp[y\phi]= & \boldsymbol{e}^{\intercal}_{y}\boldsymbol{K}\boldsymbol{v_{\phi}}=\left(\et[xy]\cos\alpha+\kappa_{y}\sin\alpha\right)\cos\phi-\et[yz]\sin\phi,\\
\kp[zr]= & \boldsymbol{e}^{\intercal}_{z}\boldsymbol{K}\boldsymbol{u}=\sin\phi\left(\et[yz]\sin\alpha+\et[xz]\cos\alpha\right)+\kappa_{z}\cos\phi,\\
\kp[z\alpha]= & \sin\phi\boldsymbol{e}^{\intercal}_{z}\boldsymbol{K}\boldsymbol{v_{\alpha}}=-\sin\phi\left(\et[xz]\sin\alpha-\et[yz]\cos\alpha\right),\\
\kp[z\phi]= & \boldsymbol{e}^{\intercal}_{z}\boldsymbol{K}\boldsymbol{v_{\phi}}=\cos\phi\left(\et[yz]\sin\alpha+\et[xz]\cos\alpha\right)-\kappa_{z}\sin\phi.
\end{aligned}
\end{equation}

Introducing the weight matrix $\boldsymbol{W}={\rm Diag\{0,0,0,1,0,0\}}$,
the estimation precision in the limit $r\rightarrow0$ can be worked
out as
\begin{equation}
\hr=\frac{\nt}{(Q^{-1})_{44}}=\frac{\hr[({\rm num})]}{\hr[({\rm den})]},
\end{equation}
where
\begin{equation}
\hr[({\rm num})]=\left(1-\epsilon^{2}\right)\nt\left(\kappa_{x}\kappa_{y}\kappa_{z}+2\et[xy]\et[yz]\et[xz]-\kappa_{x}\et[yz][2]-\et[xy][2]\kappa_{z}-\et[xz][2]\kappa_{y}\right),
\end{equation}
and
\begin{equation}
\begin{aligned}\hr[({\rm den})]= & \left(\kappa_{x}\kappa_{y}-\et[xy][2]\right)\cos^{2}\phi+\left(\kappa_{x}\kappa_{z}-\et[xz][2]\right)\sin^{2}\alpha\sin^{2}\phi+\left(\kappa_{y}\kappa_{z}-\et[yz][2]\right)\cos^{2}\alpha\sin^{2}\phi\\
 & +\left(\et[xy]\et[xz]-\kappa_{x}\et[yz]\right)\sin\alpha\sin2\phi+\left(\et[xy]\et[yz]-\et[xz]\kappa_{y}\right)\cos\alpha\sin2\phi\\
 & +\left(\et[xz]\et[yz]-\et[xy]\kappa_{z}\right)\sin2\alpha\sin^{2}\phi.
\end{aligned}
\end{equation}
Denote $\boldsymbol{u}=\left(\cos\alpha\sin\phi,\sin\alpha\sin\phi,\cos\phi\right)^{\top}$,
then the estimation precision can be further simplified as
\begin{equation}
\hr=\frac{\left(1-\epsilon^{2}\right)\nt{\rm det}\boldsymbol{K}}{\boldsymbol{u}^{\top}{\rm det}\boldsymbol{K}\boldsymbol{K}^{-1}\boldsymbol{u}}=\frac{\left(1-\epsilon^{2}\right)\nt}{\boldsymbol{u}^{\top}\boldsymbol{K}^{-1}\boldsymbol{u}}.\label{eq:hr-1}
\end{equation}
It demonstrates that the quantum-limited estimation precision of the
distance remains finite even as $r\rightarrow0$, thereby avoiding
Rayleigh's curse.

To illustrate the above results, we consider a typical Gaussian point-spread
function for the imaging system,
\begin{equation}
\begin{aligned}\psi_{i}(\boldsymbol{s})= & \frac{\exp\left[-\frac{1}{4}\left(\boldsymbol{s}-\boldsymbol{s}_{i}\right)^{{\rm \top}}\boldsymbol{\Sigma}^{-1}\left(\boldsymbol{s}-\boldsymbol{s}_{i}\right)\right]}{(2\pi)^{D/4}\left|\boldsymbol{\Sigma}\right|^{1/4}},\;i=1,2,\end{aligned}
\end{equation}
where $\boldsymbol{s}\in\mathbb{R}^{3}$ denotes the spatial coordinate
and $\boldsymbol{s}_{i}$ denotes the position of the $i$-th source,
the covariance matrix $\boldsymbol{\Sigma}$ characterizes the shape
and width of the PSF, 
\begin{equation}
\boldsymbol{\Sigma}=\left(\begin{array}{ccc}
w^{2}_{x} & \beta_{xy}w_{x}w_{y}\: & \beta_{xz}w_{x}w_{z}\\
\beta_{xy}w_{x}w_{y} & w^{2}_{y} & \beta_{yz}w_{y}w_{z}\\
\beta_{xz}w_{x}w_{z}\: & \beta_{yz}w_{y}w_{z} & w^{2}_{z}
\end{array}\right),
\end{equation}
where $\beta_{ij}\in[-1,1]$ quantifies the correlation along the
different axes, $w_{x}$, $w_{y}$ and $w_{z}$ are the widths of
the Gaussian distribution along the $x$, $y$ and $z$ axes respectively.
Then we know that
\begin{equation}
\begin{aligned}K_{ij}= & \int\partial_{i}\psi\partial_{j}\psi d^{3}\boldsymbol{s}\\
= & \int\frac{1}{4}(\boldsymbol{\Sigma}^{-1}\boldsymbol{s})_{i}\psi(\boldsymbol{\Sigma}^{-1}\boldsymbol{s}^{\intercal})_{j}\psi d^{3}\boldsymbol{s}\\
= & \frac{1}{4}\Sigma^{-1}_{ik}\Sigma^{-1}_{jl}\int\boldsymbol{s}_{k}\boldsymbol{s}^{\intercal}_{l}\left|\psi\right|^{2}d^{3}\boldsymbol{s}\\
= & \frac{1}{4}\Sigma^{-1}_{ik}\Sigma^{-1}_{jl}\Sigma_{kl}\\
= & \frac{1}{4}\Sigma^{-1}_{ij},
\end{aligned}
\end{equation}
so we have
\[
\boldsymbol{K}=\frac{1}{4}\boldsymbol{\Sigma}^{-1}.
\]
When $r\rightarrow0$, the precision $\hr$ can be further simplified
to 
\begin{equation}
\hr[\text{}]=\frac{\left(1-\epsilon^{2}\right)\nt}{4\boldsymbol{u}^{\intercal}\boldsymbol{\Sigma}\boldsymbol{u}},
\end{equation}
where
\begin{equation}
\begin{aligned}{\bf u}^{\intercal}\boldsymbol{\Sigma}{\bf u}= & w^{2}_{x}\cos^{2}\alpha\sin^{2}\phi+w^{2}_{y}\sin^{2}\alpha\sin^{2}\phi+w^{2}_{z}\cos^{2}\phi\\
 & +\beta_{yz}w_{y}w_{z}\sin\alpha\sin2\phi+\beta_{xz}w_{z}w_{x}\cos\alpha\sin2\phi\\
 & +\beta_{xy}w_{y}w_{x}\sin2\alpha\sin^{2}\phi,
\end{aligned}
\end{equation}
which remains finite in the limit of $r\rightarrow0$.

\subsection{\label{subsec:Rotation-invariance}Rotation invariance of the quantum-limited
distance precision}

Although the precision limit $\hr$ in Eq. \eqref{eq:hr-1} is expressed
in terms of the direction vector $\boldsymbol{u}$, its dependence
on the angular parameters $(\alpha,\phi)$ does not imply a dependence
on the choice of coordinate frame. The angles specify the components
of the physical displacement direction in a particular Cartesian basis,
and therefore transform together with the matrix representation of
the response tensor $\boldsymbol{K}$.

Let $\boldsymbol{R}\in\mathrm{SO}(3)$ denote an arbitrary rotation
relating two Cartesian representations of the same physical source--imaging
configuration. The momentum vector, response tensor, and displacement
direction transform as
\begin{equation}
\hat{\boldsymbol{P}}^{\prime}=\boldsymbol{R}\hat{\boldsymbol{P}},\,\boldsymbol{K}^{\prime}=\boldsymbol{R}\boldsymbol{K}\boldsymbol{R}^{\top},\,\boldsymbol{u}^{\prime}=\boldsymbol{R}\boldsymbol{u}.
\end{equation}
Since $\boldsymbol{R}^{\top}\boldsymbol{R}=\mathbb{I}$, for any rotation
matrix $\boldsymbol{R}$ one has the identity
\[
(\boldsymbol{R}\boldsymbol{K}\boldsymbol{R}^{\top})^{-1}=\boldsymbol{R}\boldsymbol{K}^{-1}\boldsymbol{R}^{\top},
\]
It follows that
\begin{equation}
\begin{aligned}\boldsymbol{u}^{\prime}{}^{\intercal}\boldsymbol{K}^{\prime}{}^{-1}\boldsymbol{u}^{\prime}= & (\boldsymbol{R}\boldsymbol{u})^{\intercal}\big(\boldsymbol{R}\boldsymbol{K}^{-1}\boldsymbol{R}^{\top}\big)(\boldsymbol{R}\boldsymbol{u})\\
= & \boldsymbol{u}^{\intercal}\boldsymbol{R}^{\intercal}\boldsymbol{R}\boldsymbol{K}^{-1}\boldsymbol{R}^{\intercal}\boldsymbol{R}\boldsymbol{u}\\
= & \boldsymbol{u}^{\intercal}\boldsymbol{K}^{-1}\boldsymbol{u},
\end{aligned}
\end{equation}
Hence the scalar contraction that enters the distance precision is
invariant under a coordinate rotation. Although the numerical values
of the angular parameters $(\alpha,\phi)$ and the matrix elements
of $\boldsymbol{K}$ may change when a different coordinate frame
is chosen, the value of $\boldsymbol{u}^{\intercal}\boldsymbol{K}^{-1}\boldsymbol{u}$,
and therefore $\mathcal{H}_{r}$, remains unchanged.

This invariance should be distinguished from a physical reorientation
of the imaging response. In a mere change of coordinates, both $\boldsymbol{u}$
and $\boldsymbol{K}$ are transformed consistently, leaving $\mathcal{H}_{r}$
unchanged. By contrast, if the source positions are fixed and the
point-spread-function response is physically reoriented, $\boldsymbol{u}$
remains fixed in the laboratory frame while the response tensor changes.
In that case, the contraction $\boldsymbol{u}^{\intercal}\boldsymbol{K}^{-1}\boldsymbol{u}$
can change, leading to a different distance precision.

\end{document}